# HeteroViT: A Versatile Single-Layer Vision Transformer Concept, Co-Designed for Distributed Real-Time Data Reduction on Scientific Detectors

Abhilasha Dave[1,*], Weijian Zheng[2], Antonino Miceli[2], Dionisio Doering[1], Ryan Herbst[1], Angelo Dragone[1],
[1] SLAC National Accelerator Laboratory, Menlo Park, CA, USA [2] Argonne National Laboratory, Lemont, IL 60439
[*] adave@slac.stanford.edu

## 1 Abstract

Next-generation X-ray detectors generate data faster than any system can affordably store or process. LCLS-II, the upgraded Linac Coherent Light Source at SLAC, produces data on the order of terabytes per second, with raw-data transfer and storage projected to be prohibitively costly, even though much of the data is not scientifically useful. This concept paper focuses on two major points. The first is versatility: a deliberately tiny, single-layer Vision Transformer (ViT) is enough to serve distinct scientific quick-evaluation tasks. We demonstrate this on two very different problems: (a) a supervised hit/miss/maybe classification on the CSPAD dataset, made to resemble ePixUHR-like detector frames, and (b) a self-supervised latent space for rare-event detection in X-ray diffraction—spanning two learning paradigms, two output types, and two detector modalities, with one small backbone. The second is hardware co-design: because the ViT's blocks are structurally uniform, the model maps cleanly onto the heterogeneous hardware already present in the LCLS detector pipeline (ASIC → FPGA → GPU) under a simple rule—one ASIC is one token—so the data is reduced progressively at each stage and a keep/discard decision is produced in real time at the edge. The two claims reinforce each other: versatility is precisely what justifies freezing the front-end in silicon, since a reusable front-end is only worth committing to hardware if it serves many tasks. We are explicit that this is a concept supported by early software analysis, not a hardware demonstration. The natural and primary next phase is the hardware implementation of this distributed pipeline. The decisive evidence still owed—an end-to-end latency budget, ASIC-feasibility of the in-sensor embedding, and the falsenegative behavior that matters for a data veto—defines that program. HeteroViT is our first step toward it.

## 2 Introduction

### 2.1 Motivation: The Data-Streaming Challenge

Modern light-source experiments have entered a regime where the bottleneck is the data itself. LCLS-I generates and stores data at gigabytes per second, while LCLS-II produces data at terabytes per second. Streaming raw data every microsecond would overwhelm the existing network, storage, and processing infrastructure, and archiving this volume is estimated to cost on the order of billions of dollars per year—compounded by the fact that a large fraction of recorded frames carry no scientific value. Our goal is to autonomously select valid scientific data and discard the rest in real time. Rather than recording everything and sorting it later, we run machine-learning classification and data extraction at the edge, directly on or near the detector, to separate meaningful events from background and extract insight as close to the sensor as possible before the data ever reaches central storage.

## 2.2 The Centralized Inference Problem

The intuitive deployment target to run these ML models are GPUs, which offers mature libraries and tooling. But tracing the data's journey from ASICs to FPGAs that stitches signals from many sensors, then potentially a GPU [13] exposes the problem. GPUs are built for batch processing [5, 9], while detectors stream one frame at a time. Feeding a GPU efficiently means buffering many frames and shipping them off, which saturates the network (pushing raw images across the very link we are trying to relieve) and adds latency (by the time the data completes its round trip, the moment has passed). At the detector edge, decisions are needed on the timescale of microseconds per frame: is this frame interesting (keep it) or junk (discard it)? Batching overwhelms the network and storage, while a batch size of one leaves the GPU slow and power-hungry. That fast triage belongs at the edge, not in a distant data center.

## 2.3 Distributed ML Across Heterogeneous Hardware

Pushing the model toward the front end runs into a hard limit: an entire model cannot fit on a single FPGA [2] or ASIC, which are tightly constrained in memory, compute, and power. Our proposal is therefore to split the model across the multiple devices already present in the pipeline, spreading the workload across the heterogeneous hardware that exists. The goal is explicitly not to replace the detector pipeline but to enhance it—layering smarter decision-making onto the existing ASIC, FPGA, and GPU stages. As data rates climb, the question is no longer whether to use AI but where to run it; our answer is to place the intelligence at the edge, inside the detector itself.

## 2.4 Why a Vision Transformer: Modularity, Not Superiority

Deploying modern AI at the detector edge runs into a tension in timescales: hardware lifecycles are measured in years (the reality of FPGA and ASIC design) while ML algorithms evolve in months. Hard-wiring today's best algorithm into silicon risks obsolescence before the detector turns on [3]. The property we actually need is modularity: a front-end that can be fixed in silicon and reused unchanged, paired with a small reconfigurable tail that absorbs algorithm dependability. For example the difficulty with convolutional networks is not that they cannot be partitioned, but that their layers are heterogeneous in shape and operation, so splitting one across ASIC → FPGA → GPU means hand-managing inter-stage data movement, timing, synchronization, buffering, and protocol bridging at every boundary. Any architectural tweak ripples through the whole chain, forcing re-verification and re-timing. In principle a frozen CNN feature extractor with a reconfigurable head could provide similar modularity; the real question is which architecture makes the fixed/reconfigurable boundary cleanest and most stable across applications. In this research we've opted for narrow approach singular architecture of transformer neural networks: because every transformer block has an identical shape and compute profile [11, 4], the partition is unusually clean. A single backbone is mapped once and sliced along well-defined interfaces embedding and early operations fixed on ASIC/FPGA, and only the task head, on the reconfigurable GPU, changing between applications. We are not claiming a transformer is intrinsically more powerful than any other ML models for this task; at the scale we use, the model is deliberately tiny. The claim is about deployability: homogeneous blocks give a stable, repeatable hardware.

# 3 The Miniature HeteroViT: System Architecture and HardwareAlgorithm Co-Design

## 3.1 The ePixUHR data-reduction pipeline

The target system is the end-to-end data-reduction pipeline of the LCLS-II ePixUHR detector, a hierarchical architecture that moves from raw sensor capture to offline analysis Figure 1. The frontend is a dense array of ASICs [10] in a 12×12 grid (144 ASICs total), each producing a 192×168-pixel frame (roughly 4.6 megapixels per full frame). The mid-level readout uses FPGA boards in a 4×6 configuration, each aggregating data from six ASICs.

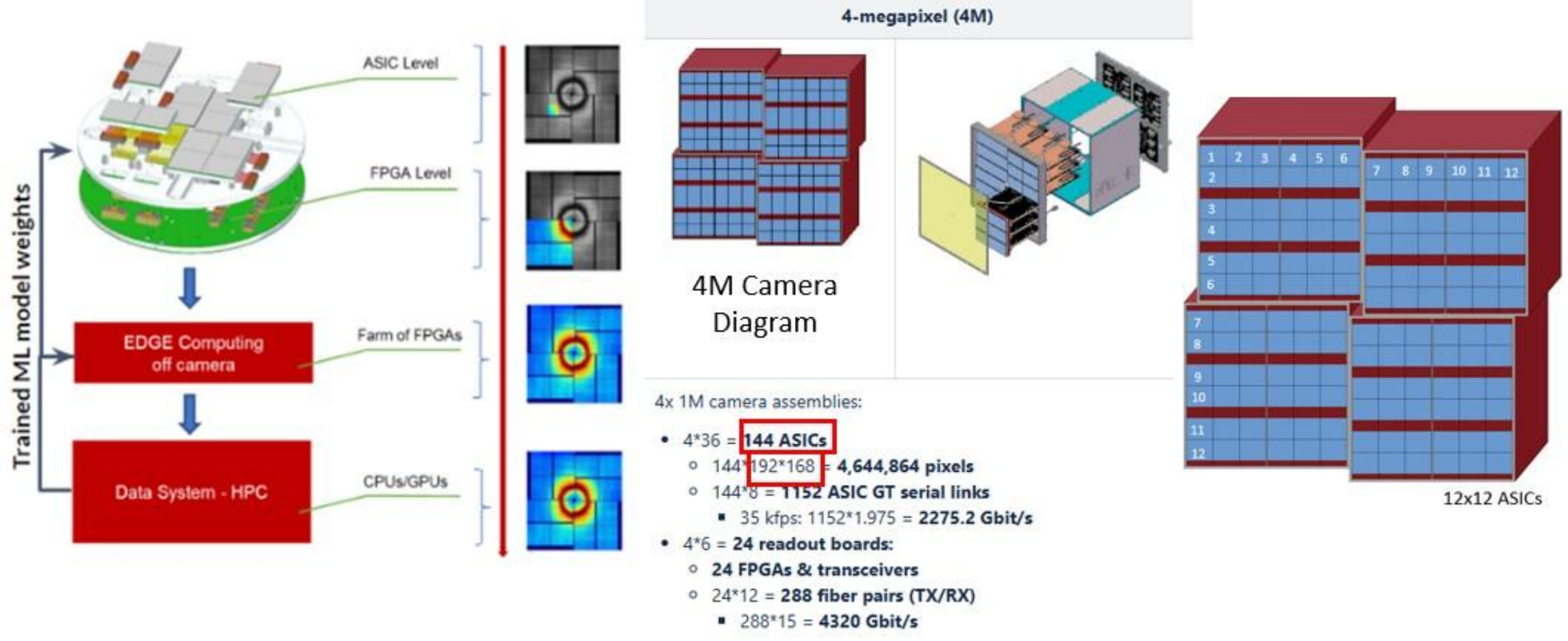

Figure 1: The ePixUHR hardware stages targeted by HeteroViT 144 ASICs (12×12), an FPGA readout farm, and a CPU/GPU back-end. Form the detector data pipeline: ASIC to FPGA to GPU/CPU [12]

## 3.2 One ASIC, one token — the core mapping

The Vision Transformer splits an image into patches, treats each patch as a token, embeds it, and lets self-attention learn global relationships between tokens [4]. HeteroViT carries this one decisive step further: each ASIC is treated as a single token. The strength of this mapping is that it is not forced the hardware's natural boundary and the architecture's natural boundary both are genuinely coincide. Each ASIC is a physically independent unit producing its own region of the frame, and a token is a unit processed independently right up until attention. Lining the two up means the physical detector layout maps directly onto the model's sequence dimension.

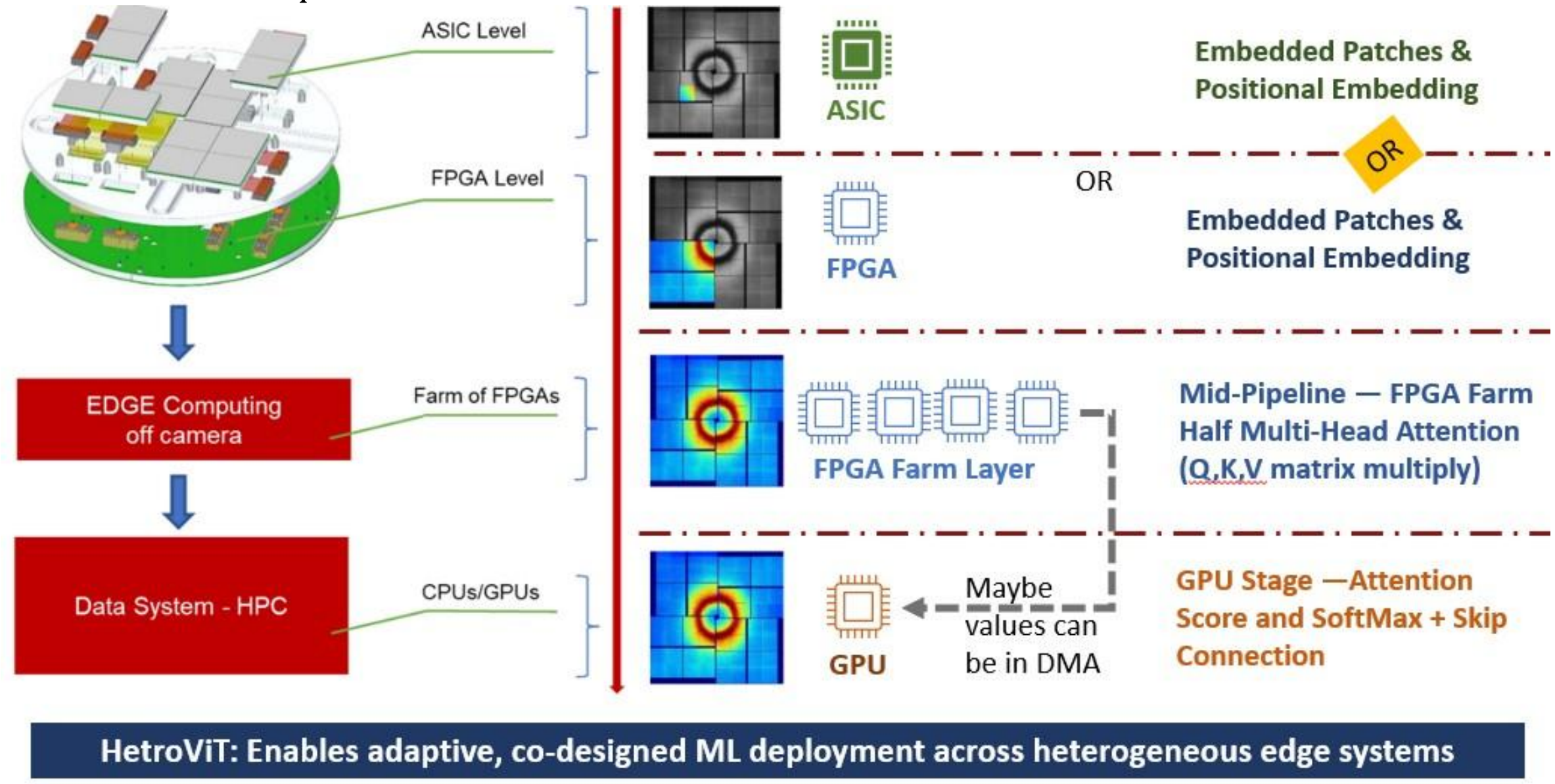

Figure 2: Mapping the Vision Transformer onto the detector hierarchy: patch and positional embedding on the ASIC (or FPGA) layer, partial multi-head attention on the FPGA farm, and attention score, softmax, and the task head on the GPU stage

This mapping carries one honest trade-off worth stating as a design decision rather than leaving implicit: collapsing an ASIC's 192×168 pixels into a single token discards spatial detail within an ASIC.

This is by design — the granularity is well matched to real-time data reduction and coarse hit/miss triage at the front of the pipeline, where the goal is to decide quickly what is worth keeping. It might, however, be too blunt for downstream science that requires localizing a feature inside an individual ASIC's field of view.

## 3.3 Distributed Computing and Layer-wise Data Reduction

The mapping in Figure 2 leads to a natural distributed-computing scheme: each operation runs on the hardware best suited to it, and the data volume shrinks at every stage. Operations that act on one token at a time — embedding and the Q/K/V projection — are independent and can run in parallel, so they live on the distributed ASIC and FPGA front-end. The one operation that couples all tokens together — the attention score and softmax — runs on the GPU back-end, where every token is gathered in one place. The result is a pipeline that progressively reduces data as it flows downstream, ending in a real-time keep-or-discard decision.

### 3.3.1 How much reduction does the mapping itself buy?

The data-reduction effect is a property of the mapping, not of any particular embedding size. To make this concrete, Table 1 reports the reduction at the ASIC layer using the three standard ViT embedding dimensions from the literature [4, 14]: ViT-Base (D=768), ViT-Large (D=1,024), and ViT-Huge (D=1,280).

| VIT VARIANT | D | PER-ASIC TOKEN SIZE | REDUCTION AT ASIC LAYER |
|---|---|---|---|
| ViT-Base | 768 | 3.0 KB | ~95.2% |
| ViT-Large | 1,024 | 4.0 KB | ~93.6% |
| ViT-Huge | 1,280 | 5.0 KB | ~92.1% |

Table 1: Embedding Reduction. Assuming 16-bit raw pixels and float32 embeddings; raw frame per ASIC = 192 × 168 × 2 B ≈ 63 KB.

Even if we take ViT-Huge embedding dimention with full float32, the one-ASIC-one-token mapping alone removes more than 90% of the per-ASIC data volume before any attention computation takes place. The FPGA stage further structures this output into Q, K, V projections, after which only those projections, not the raw frames, need to traverse the link to the GPU back-end.

**A note on the case studies in this paper:** The two case studies presented in section 4 section 5 are preliminary proofs-of-concept. They use a deliberately small embedding space to keep the demonstration tractable, which pushes the per-ASIC reduction to roughly 99.85%. This number should not be read as the operating point of a deployed system; it reflects the embedding choice in these particular studies. The point we wish to convey is the one made by Table 1: the architectural mapping yields substantial, progressive data reduction across the full range of standard ViT embedding sizes. Scaling to ViT-Base/Large/Huge dimensions, and characterizing the resulting accuracy/throughput trade-off, is left to future work.

### 3.3.2 ASIC stage: in-sensor embedding

Each ASIC's 192×168 = 32,256-pixel output is flattened, projected by a dense embedding layer into a compact space of dimension D, and combined with a positional encoding. For the preliminary ePixUHR model presented here, D=25: a dense layer with 25 neurons compresses the 32,256 pixels into a 25-dimensional token at a cost of roughly 0.8 M MAC operations, so only 25 values per ASIC are sent downstream instead of all 32,256 pixels — about 99.85% data reduction per ASIC (cf. Table 1 for the corresponding figures at standard ViT embedding sizes).

It is worth being precise about what drives this number. The reduction comes from computing the first layer at the edge; the same effect would follow from any first layer that compresses the frame. The transformer is not what makes the reduction large it is what makes the cut clean, because that first layer is a simple per-token dense operation with a fixed, hardware-friendly shape. The decisive question is therefore feasibility, not arithmetic: whether a 32,256→$D$ dense layer fits the ePixUHR ASIC's power and area budget at the detector's frame rate, and in what numerical precision.

#### 3.3.3 FPGA stage partial multi-head attention

Each FPGA receives the six embedded tokens from its six ASICs and matrix-multiplies them with the Key, Query, and Value weight matrices, one set per attention head, emitting the partial attention projections. For the preliminary configuration used here (5 heads, D=25), this amounts to roughly 99.8% data reduction per FPGA. These percentages are specific to this configuration; they vary with embedding dimension D, head count, and precision, and the corresponding numbers at standard ViT sizes follow the same trend reported in Table 1. Because the FPGA is reconfigurable, a future change to the number of heads — or to D itself — requires modifying only this stage, leaving the ASIC untouched.

#### 3.3.4 GPU stage global attention and decision

The remaining computations the global attention score, softmax, skip connection, and task-specific MLP head run on the GPU, which produces the final objective (classification or embedding). Data moves directly from the FPGAs to the GPU via Direct Memory Access (DMA) transactions, without intermediate batch storage.

It is worth being explicit about which part of attention is being distributed and which is not. The Q/K/V projections on the FPGA are per-token and embarrassingly parallel, so they distribute cleanly. The genuinely coupled operation the all-to-all $QK^T$ score and softmax across every token is global and is deliberately kept on the GPU. The scheme therefore distributes the cheap, parallel work and centralizes the hard part.

# 4 Case Study I: Edge Classification

Because ePixUHR data is not yet widely available, we use CSPAD data reformatted to ePixUHR-like 4-megapixel dimensions as a close stand-in. The proposed single-layer ViT is run in software, with the hardware partitioning simulated rather than executed, and tasked with sorting frames into hit, miss, and maybe categories. The model reaches 85.5% three-class accuracy. We read this as a feasibility signal a deliberately small model retains usable capacity for first-pass triage and no more than that.

For context [8] applied a CNN-based screening tool to the closely related LG36 CSPAD dataset and reported 91% classification accuracy on the same hit/miss problem Our 85.5% therefore sits a few points below an established CNN baseline on comparable data, while operating under a far tighter computational budget.A single transformer layer with D=25, designed to be partitioned across ASIC, FPGA, and GPU stages rather than run as a monolithic offline classifier. The point of comparison is not that the tiny ViT wins on accuracy, but that it remains in the same ballpark while being deployable in-pipeline. One important thing to note here is the CNN comparison is suggestive but not a controlled benchmark. A fair head-to-head requires running both models on identical splits, with matched preprocessing and matched parameter budgets.

# 5 Case Study II: Self-Supervised Representations

The second study addresses a different problem from high-energy X-ray diffraction, which non-destructively maps the 3D microstructure of bulk metallic polycrystalline materials, typically under thermo-mechanical loading to capture microstructural evolution. As in the detector case, massive data volumes and slow traditional pipelines limit insight extraction; detecting rare events such as rotations, phase transitions, or cracks is especially hard.

This study probes a different axis of evidence than Case Study I. It does not test hardware distribution; it tests model capacity and versatility — whether the same small single-layer ViT, now trained without labels, can learn an embedding rich enough to resolve a real physics transition.

## 5.1 Self-supervised learning with BYOL

Because scientific data is rarely well labeled, we train the ViT using BYOL (Bootstrap Your Own Latent), which learns patch embeddings without labels [6]. Two augmented views of the same patch feed an online network (encoder + projector + predictor) and a target network (encoder + projector); the online network predicts the target's representation of the other view, with a cosine-distance loss. Collapse is avoided without negative samples because the target network is not trained by backpropagation but updated as an exponential moving average (EMA) of the online encoder.

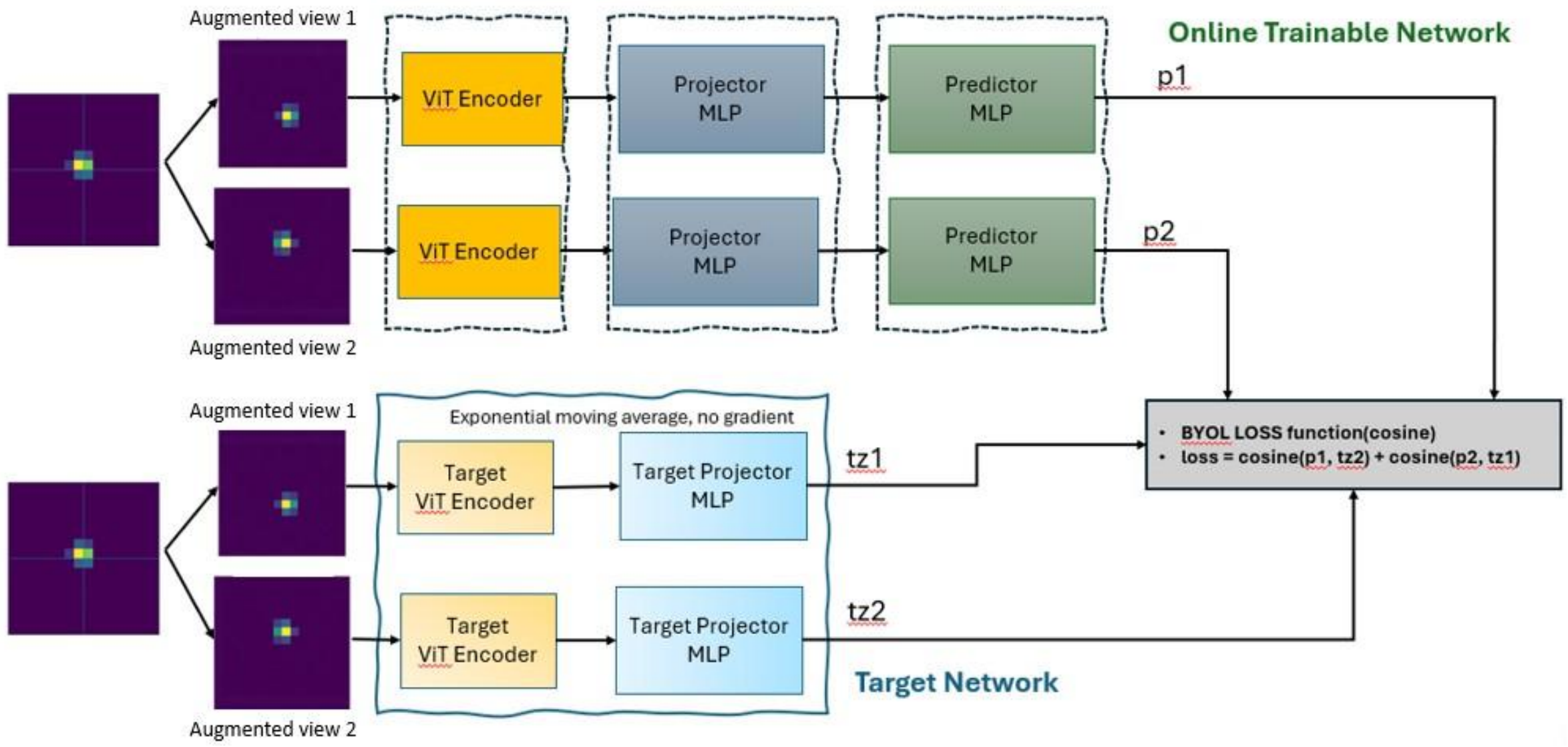

Figure 3: The BYOL-ViT training scheme. The online network learns to predict the target network's projected representation of a second augmented view; the target weights are an EMA of the online encoder.

### 5.1.1 Self-supervised learning with BYOL

Because scientific data is rarely well labeled, we train the ViT using BYOL (Bootstrap Your Own Latent), which learns patch embeddings without labels. As shown Figure 3 two augmented views of the same patch feed an online network (encoder + projector + predictor) and a target network (encoder + projector); the online network predicts the target's representation of the other view, with a cosine-distance loss. Collapse is avoided without negative samples because the target network is not trained by backpropagation but updated as an exponential moving average (EMA) of the online encoder.

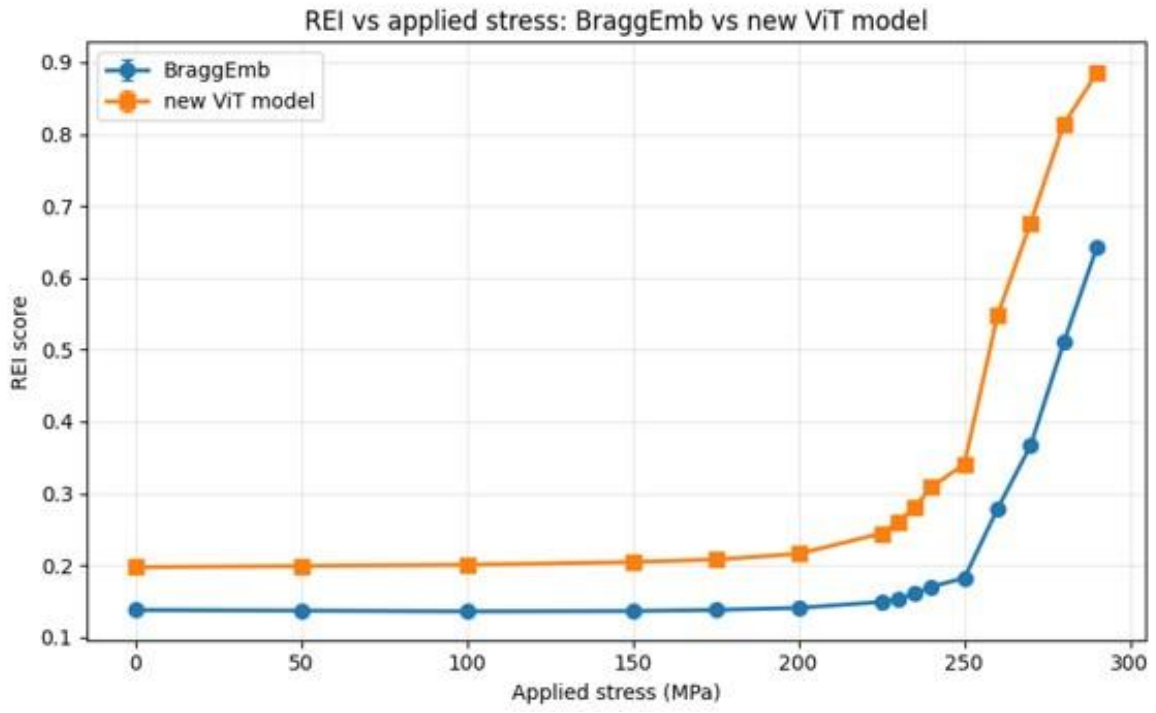

Figure 4: Rare-event indicator (REI) versus applied stress for the established BraggEmb model and the single-layer ViT. Both capture the sharp rise near the elastic-to-plastic transition ( 250 MPa). Reference: [15]

**Validation:** The relevant physics is the stress–strain response. In the elastic regime, the curve is linear and the metal springs back; at the yield point the "knee" atoms slide permanently along crystal slip planes and the metal deforms plastically. This transition is the rare but critical event to detect. As shown in Figure 4 on this dataset, the embedding of our single-layer BYOL-ViT tracks the elastic-to-plastic transition with a rare-event indicator comparable to the established BraggEmb model shown in this work [15]. The scope of the claim is that a lightweight, label-free ViT can match an established CNN baseline on this one rare-event task a useful capacity result.

## 6 Disscussion

The two case studies are not separate demonstrations; together they make a single point. The same tiny single-layer backbone handled two genuinely different problems: a supervised classification that emits a decision, and a self-supervised representation that emits an embedding. Two learning paradigms, two output types, and two detector modalities (area-detector frames and flat-panel diffraction patches). Each illustrates a distinct flavor of edge evaluation classification closes the loop with a direct store/veto decision, while BYOL provides label-free representations for the common case where labels do not exist.

This is where the architecture and the hardware reinforce each other. Freezing a front-end in an ASIC is only sensible if that front-end is reusable across tasks, and the two studies are preliminary evidence that one fixed embedding-and-attention front-end can serve in that role. Versatility is therefore not a side result; it is the argument for the hardware modularity. The versatility result says why a fixed front-end is worth committing to silicon; the co-design says how one would commit it. Read together, the two halves form a co-design in which the architecture choice and the hardware choice each justify the other.

**We keep the scope modest:** Two tasks are encouraging evidence of versatility, not proof of a general backbone. Extending the study to additional scientific datasets and task types is the natural next step toward strengthening this claim.

## 7 Conclusion

We presented HeteroViT, a hardware–algorithm co-design that maps a small Vision Transformer onto the distributed compute hierarchy of the LCLS-II ePixUHR detector. The contribution is twofold: a hardware-topology-aware token mapping — one ASIC, one token that performs in-sensor dimensionality reduction, and preliminary evidence that a single small ViT backbone is versatile enough to serve distinct scientific quick-evaluation tasks.

The two case studies make a single point: the same tiny backbone handled both a supervised classification 85.5% three-class accuracy on ePixUHR-like data, against 91% from the CNN baseline of [8], and a self-supervised representation that matched the BraggEmb baseline on a rare-event task [15]. Two learning paradigms, two output types, two detector modalities one fixed front-end. This is the central argument of the paper: versatility is what makes committing such a backbone to silicon worthwhile, and the co-design is what makes the commitment realizable. The architecture choice and the hardware choice each justify the other.

We keep the scope modest. Two tasks are encouraging evidence, not proof of a general backbone, and the studies are software based rather than hardware demonstrations. The engineering work required to validate the concept on the ePixUHR detector chain is described in Future work.

## 8 Future Work

The concept must now be validated on hardware. Our next phase focuses on developing the supporting libraries and producing a prototype implementation on the ePixUHR detector chain. Specifically:

- **Library development** Implement the per-stage ViT operations — embedding, Q/K/V projection, and attention using the SLAC Neural Network Libraries (SNL) [7, 2, 1], so that the same model description can be deployed across ASIC, FPGA, and GPU stages.

- **ASIC feasibility of the embedding** Determine whether the in-sensor dense layer fits the ePixUHR power and area budget at full frame rate, and at what numerical precision. This is the decisive engineering question and currently dictates the ASIC–FPGA placement of the embedding.

- **End-to-end latency** Quantify cumulative latency across ASIC embedding, FPGA Q/K/V projection, GPU attention and softmax, and the FPGA→GPU DMA transport.
- **Generalization across tasks** Survey scientific datasets and quick-analysis tasks across LCLS-II beamlines to test whether a single fixed ViT front-end can be reused across them in silicon.

# 9 Acknowledgments

The work of the authors is supported by the U.S. Department of Energy under Contract No. DEAC02-76SF00515, SLAC FWP 101264.

# 10 Generative AI Statement

The author(s) declare that Gen AI was used in the creation of this manuscript. We acknowledge the assistance of the large language model in refining the language and enhancing the readability of this paper.

# References

[1] Abhilasha Dave et al. "FPGA-Accelerated Real-Time Diagnostics at DIII-D Using the SLAC Neural Network Library for ML Inference". In: *arXiv preprint arXiv:2604.26042* (2026).

[2] Abhilasha Dave et al. "FPGA-accelerated SpeckleNN with SNL for real-time X-ray single-particle imaging". In: *Frontiers in High Performance Computing* 3 (2025), p. 1520151.

[3] Allison McCarn Deiana et al. "Applications and techniques for fast machine learning in science". In: *Frontiers in big Data* 5 (2022), p. 787421.

[4] Alexey Dosovitskiy et al. "An image is worth 16x16 words: Transformers for image recognition at scale". In: *arXiv preprint arXiv:2010.11929* (2020).

[5] Yanjie Gao et al. "An empirical study on low gpu utilization of deep learning jobs". In: *Proceedings of the IEEE/ACM 46th International Conference on Software Engineering*. 2024, pp. 1– 13.

[6] Jean-Bastien Grill et al. "Bootstrap your own latent-a new approach to self-supervised learning". In: *Advances in neural information processing systems* 33 (2020), pp. 21271–21284.

[7] R. Herbst et al. "Implementation of a framework for deploying AI inference engines in FPGAs". In: *Smoky Mountains Computational Sciences and Engineering Conference*. Springer, 2022, pp. 120–134.

[8] T-W Ke et al. "A convolutional neural network-based screening tool for X-ray serial crystallography". In: *Synchrotron Radiation* 25.3 (2018), pp. 655–670.

[9] Pol G Recasens et al. "Mind the memory gap: Unveiling gpu bottlenecks in large-batch llm inference". In: *2025 IEEE 18th International Conference on Cloud Computing (CLOUD)*. IEEE. 2025, pp. 277–287.

[10] H Sandberg et al. "GT Readout—A development platform for 1 MHz frame-rate detectors at LCLS-II". In: *Journal of Instrumentation* 20.08 (2025), P08019.

[11] Mohammad Shoeybi et al. "Megatron-lm: Training multi-billion parameter language models using model parallelism". In: *arXiv preprint arXiv:1909.08053* (2019).

[12] SLAC National Accelerator Laboratory. *AUREIS: Adaptive Ultra-fast Energy-efficient Intelligent Sensing*. https://lcls.slac.stanford.edu/depts/data-systems/projects/aureis. Accessed: June 8, 2026. 2026.

[13] Jana Thayer et al. "Massive scale data analytics at LCLS-II". In: *EPJ Web of Conferences*. Vol. 295. EDP Sciences. 2024, p. 13002.

[14] Ashish Vaswani et al. “Attention is all you need”. In: *Advances in neural information processing systems* 30 (2017).

[15] Weijian Zheng et al. “Rapid detection of rare events from *in situ* X-ray diffraction data using machine learning”. In: *Journal of Applied Crystallography* 57.4 (Aug. 2024). doi: 10.1107/S160057672400517X. url: https://doi.org/10.1107/S160057672400517X.